\documentstyle[preprint,aps]{revtex}                 
\begin{document}
\pagestyle{empty}                              
\preprint{
\font\fortssbx=cmssbx10 scaled \magstep2
}
\draft
%
%
\title{
MORE ON THE SU(2) DECONFINEMENT TRANSITION IN THE MIXED ACTION
}
\vfill
\author{Rajiv V. Gavai\footnote{Email:gavai@mayur.tifr.res.in}}
\address{
Theoretical Physics Group, Tata Institute of Fundamental Research \\
Homi Bhabha Road, Mumbai 400005, India
}
\author{Manu Mathur\footnote{Permanent address: S. N. Bose National Centre for 
Basic Sciences, JD Block, Sector III, Slat Lake, Calcutta 700064, India. 
Email:manu@hpth.difi.unipi.it}}
\address{
Dipartimento di Fisica dell' Universit\^a and I.N.F.N \\
Piazza Torricelli 2 Pisa-56100, Italy
}
%
%
\vfill

\maketitle
\begin{abstract}
We examine certain issues related to the
universality  of the SU(2) lattice gauge theory at
non-zero temperatures. Using Monte Carlo simulations and strong coupling
expansions, we study the behavior of the deconfinement transition in
an extended coupling plane $(\beta,\beta_A)$ around the tricritical
point where the deconfinement transition changes from second to first
order. Our numerical results on $N_\tau$ =2,4,6,8 lattices show that the
tricritical point first moves down towards the Wilson axis and and then
moves slowly upwards, if at all, as the lattice spacing is reduced. Lattices 
with very large $N_\tau$ seem to be therefore necessary for the mixed action
to exhibit the critical exponents of the three dimensional
Ising model for positive values of the adjoint coupling.
\end{abstract}
%
%
%
%
\pagestyle{plain}
 
\begin{center}
1. \bf INTRODUCTION \\
\end{center}
\bigskip
Confinement of the non-abelian color degrees of freedom has been a 
challenging problem ever since gauge theories were formulated for quark-gluon 
interactions. In the beginning, our understanding of such gauge theories 
followed mainly from perturbation theory. Due to the property of asymptotic 
freedom, the perturbative approximations are valid for short distance 
phenomenon but they are inadequate to explore long distance physics like 
confinement of quarks. A strong evidence for, and a much better insight of, 
the color confinement mechanism  in non-abelian gauge theories
has been provided by analytical computations and Monte Carlo simulations 
of quantum field theories with a non-perturbative lattice regularisation.
The simplest of such theory is described by the $SU(2)$ 
Wilson action \cite{Wil}. It was expected 
that non-abelian gauge theories in general do not have any phase transitions 
separating strong and weak coupling regimes. Therefore confinement, 
explicitly shown on the lattice in the strong coupling region, should 
persist also in the continuum limit. However, it was later found \cite{HalSch} 
that certain non-abelian lattice gauge theories 
(e.g. $SO(3)$, $SU(4)$, $SU(5)$), with Wilson form of action 
undergo bulk transitions separating strong confining 
region from the weak coupling region where the continuum limit of the theory 
exists. Bhanot and Creutz \cite{BhaCre}, extending the form of the
action proposed by Wilson, showed that this apparent loss of 
confinement can be attributed to lattice artifacts associated with the choice
of action, namely the so-called bulk phase transitions.

Subsequent to the work of Bhanot and Creutz to characterize the 
bulk transitions in the extended coupling plane, important reasons 
for further exploration of this action in the past have been to study the 
basic mechanism of confinement itself \cite{CanHalSch} and to find out 
the origin of these bulk transitions \cite{AlbFly,OgiHor,DasHel}. It 
has been a common folklore that the abrupt change from the strong coupling 
region to the scaling region for the Wilson action is due to the proximity 
of the critical point at the end of the first order line CD in Fig. 1,
where the phase diagram obtained in Ref. \cite{BhaCre} is shown,  and that
a ``smoother'' continuum limit may be obtained by going to negative $\beta_A$,
the additional coupling for this action.  Due to the theoretical
expectations of the role of $SO(3) (SU(N)/Z_N)$
monopoles in $SU(2) (SU(N))$ confinement \cite{Tom1}, the above model 
is tailor made to study the interplay of these topological degrees of 
freedom and their role in confinement between $SU(2)(SU(N))$ and $SO(3)
(SU(N)/Z_N)$ lattice gauge theories.
In fact, the plaquette susceptibility peak in the cross-over 
region in $SU(2)$ lattice gauge theory and the bulk transition line $BCD$
have been attributed to the underlying $SO(3)$ theory and its first 
order transition \cite{Tom2,BroKes}.  These issues can be also 
analyzed and tested by exploring the mixed action at non-zero 
temperatures and will be further discussed after presentation of 
our results.

The rich phase diagram associated with the mixed action, shown in Fig. 1 
by solid and dashed lines, was established mainly by Monte Carlo simulations on 
relatively small lattices \cite{BhaCre} $(4^4$-$5^4)$ with periodic boundary 
conditions. Since these small lattices were also at finite temperature, the 
phase diagram is incomplete in the absence of the deconfinement 
transition line. Along the $\beta_A =0.0$ axis, several finite temperature 
investigations have shown the presence of a second order deconfinement phase 
transition. Its critical temperature has been shown\cite{FinHelKar} to 
exhibit asymptotic scaling and its critical exponents have been 
shown\cite{EngFin} to be in very good agreement with those of the three 
dimensional Ising model.  Effective field theory arguments for the order
parameter were used by Svetitsky and Yaffe\cite{SveYaf} to conjecture
the finite temperature SU(2) gauge theory and the three dimensional
Ising model to be in the same universality class.  The verification of
this universality conjecture thus strengthened our analytical understanding 
of the deconfinement phase transition.  Our work \cite{us1,us2} on the 
extended action at non-zero temperatures began with the motivation to locate 
the line of deconfinement transition in the couplings plane $(\beta, \beta_A)$. 
Our simulations yielded the following surprising results:

\begin{enumerate} 

\item[{a]}] The transition remained second order in agreement with the 
universality conjectured exponents up to $\beta_A \approx 1.0$ but it became 
definitely first order for large enough $\beta_A (\ge 1.4)$. 

\item[{b]}] There was no evidence of a second separate transition at
larger $\beta_A$, as would be suggested by the claim of Ref. \cite{BhaCre}
of a bulk transition there.  For $N_\tau=4$
lattices the line of deconfinement transition was coincident with the
line of bulk transitions of Ref. \cite{BhaCre} but for $N_\tau=2$ there
were {\it no} symptoms of any transition at those locations.  The line of
deconfinement phase transition, on the other hand, did move to smaller
$\beta$ for all $\beta_A$, as $N_\tau$ changed from 4 to 2.

\end{enumerate}

While the details of our analysis and results can be found in the works
cited above, the key findings which lead us to these conclusions were
following:
 
\begin{enumerate}

\item[{a]}] The deconfinement order parameter, $\langle L \rangle$ (see next 
section for definition),  acquired nonzero large value at the only transition 
found on all lattices studied (i.e $N_\tau$=2,4,6,8) and
showed clear co-existence of both phases at the transition point for
larger $\beta_A$.

\item[{b]}] The same critical exponent which established the transition
to be in the Ising model universality class for $0.0 \le \beta_A \le 1.0$ 
became equal to the space dimensionality, as a first order deconfinement
phase transition would have, for larger $\beta_A$.

\item[{c]}] The plaquette susceptibility showed a decrease at $\beta_A=1.1$ 
when the lattice 4-volume was increased by a factor of 16; it should 
diverge, i.e., increase 16-fold, if there were a first order bulk phase
transition at $\beta_A=1.1$.

\end{enumerate}

\noindent
The plaquette susceptibility results above are very similar to those of Ref.
\cite{EngSch} who too found a decrease in it while increasing the
lattice volume by a factor of 16.  On increasing the lattice size
further, no further finite size dependence was found, leading to a
conclusion that the finite size effects on smaller lattices are due to
finite temperature effects.  While larger lattices will be needed in our
case too to see if a similar conclusion is reached, it has to be
emphasized that conclusions based on finite size scaling usually do
assume that the lattices are large enough for the scaling to set in.
Thus a distinguishing feature between the bulk and deconfinement phase
transitions, {\it i.~e.} the finite size scaling behavior of the coupling at 
which the transition takes place with the temporal size of the lattice
which leads one to expect the bulk transitions to move much less compared
to the deconfinement transitions, is not necessarily useful here 
since it is not clear how big lattices are
needed for this behavior to set in at various values of
$\beta_A$.  We have therefore relied heavily on the order parameter
$\langle L \rangle$ to label a transition as a deconfinement phase
transition, as mentioned in a] above.

Recently, the above surprising results showing the change 
in the order of the deconfinement transition and the absence of the bulk 
transition were confirmed \cite{Ste} for another variant of the SU(2) action
with a Villain form for the adjoint SO(3) part. We will later comment
more on the above action. 
Taken together, these results pose many questions about the continuum limit 
of the deconfinement phase transition and about the existence of
separate bulk phase transitions.  The foremost amongst them is about an
apparent {\it qualitative} violation of the universality\footnote{Note that 
this universality, which results from the freedom of choice of the lattice 
type and action, is different from the finite temperature universality 
of critical exponents discussed earlier.}, since an apparently irrelevant 
coupling seems to change the order of the deconfinement phase
transition.  The early simulations of the $SU(2)$ lattice 
gauge theory are known to have yielded {\it quantitative} violations of 
universality\cite{BhaDas}.  However, attributing them to the ignored higher 
orders in $g_u^2$, it has
been shown\cite{MuSch,Gav,GavKarSat} that dimensionless ratios of physical  
quantities have much weaker such violations.  Indeed, one can hope that
these violations will smoothly disappear under the error bars of the
simulations as the cut-off becomes smaller.   This will obviously not be
the case for any qualitative violations of universality.  Of course, the
region of couplings where universal results are obtained may have such an
irregular shape that still larger lattices are needed to obtain 
universal results. It is not clear in that case,
however, what the universal result would be.  Clearly, if a universal result
exists in the $a \rightarrow 0$ limit, then  the tricritical point
T, where the deconfinement phase transition changes order, must not
appear on any renormalisation group flow lines to the $g_u=0.0$ critical point
and must thus be invisible in the continuum limit. If the point T moves up to
large positive $\beta_A$ with increasing $N_\tau$ (and decreasing 
lattice spacing $a$), then the SU(2) deconfinement phase transition
could still be of second order in continuum limit with Ising model exponents.
This would be so irrespective of $\beta_A$ used for simulations. If, on
the other hand, the point T moves to large negative $\beta_A$, the
universality with Ising model will be lost and the transition will be
first order, again irrespective of $\beta_A$ used in simulations.  

It may be argued that the presence of a line of bulk phase
transitions and its end point will strongly modify the approach to
continuum limit and thus large lattices are mandatory for seeing the
universal physics at large $\beta_A$.  It needs to be noted therefore
that the bulk line in question was established {\it only} in numerical
simulations on small, $4^4$-$5^4$ lattices.   
A recent simulation\cite{gavbulk} at $\beta_A$ = 1.25 on larger $N^4$ 
lattices, with $N$ = 6, 8, 10, 12 and 16, found 1) a linear decrease in the 
average discontinuity in plaquette, $\Delta P$, with $N$ and 2) a plaquette
susceptibility exponent of 2.09$\pm$ 0.31 in contrast to the expected
value of 4 for a first order bulk phase transition.  This suggests that
the end point of the bulk line is at $\beta_A > 1.25$. 
This is explicitly shown in Fig. 1 by drawing solid and dashed bulk lines 
above and below $\beta_A=1.25$ respectively. 
While the result of Ref.\cite{gavbulk} does explain the above mentioned results 
on plaquette susceptibility, the mystery of the apparent coincidence of the two 
different transition lines still remains for larger $\beta_A$.
As we will show below, the deconfinement phase transition for $N_\tau$ = 4 
lattices, as identified by the order parameter, $\langle |L| \rangle$,
turns first order already at $\beta_A=1.25$, suggesting that the bulk line 
or its end point are unlikely sources of this change.

In this paper, we address the issue of the trajectory of point T with
decreasing lattice spacing $a$, after defining in the next section the
action we investigate and the observables we use along with their
scaling laws.   A simple strong coupling calculation is presented 
in Section 3, which suggests that the point T moves up in the plane to 
infinity. However, in our numerical simulations, described in Section 4, 
we find that it moves down on going from $N_\tau$ =2 to 4. 
On increasing $N_\tau$ further to 6 and then 8, we 
observe a very small upward movement by comparing the relative shapes of 
the Polyakov loop histograms. The last section  contains
a brief summary of our results and their discussion.

\bigskip
\begin{center}
2.  THE MODEL AND THE OBSERVABLES \\
\end{center}
\bigskip

The lattice action is constrained only by a) the gauge
invariance and b) the limit of zero lattice spacing which must coincide
with the continuum form of the action.  Infinitely many different forms
satisfying these criteria can be written down.  Bhanot and Creutz extended the
Wilson action to a form described by the action,

\begin{eqnarray}
S = \sum_P \left( \beta \left(1 - {1\over2} Tr_F U_P \right) + 
         \beta_A \left(1 - {1\over3} Tr_A U_P \right) \right)~~~~~.
\label{ea}
\end{eqnarray}

Here $U_P$ denotes the directed product of the basic link
variables which describe the gauge fields, $U_\mu(x)$, around an
elementary plaquette $P$. $F$ and $A$ denote that the respective traces are
evaluated in fundamental and adjoint representations respectively.  
Comparing the naive classical continuum limit of eq. (\ref{ea}) 
with the standard $SU(2)$ Yang-Mills action, one obtains
\begin{eqnarray}
{1 \over g^2_u} = {\beta \over 4} + {2 \beta_A \over 3}~~.~~
\label{gu} 
\end{eqnarray}
Here $g_u$ is the bare coupling constant of the continuum
theory.  Introducing another coupling $\theta$, defined by
$\tan\theta = \beta_A/\beta$,  the asymptotic scaling relation 
\cite{KorAlt} for this action is
\begin{eqnarray}
a = {1 \over \Lambda (\theta)} \exp\left[- {1 \over 2\beta_0 g^2_u}\right]
\left[\beta_0 g^2_u\right]^{-\beta_1 \over 2\beta^2_0} ~~~,~~
\label{ga}
\end{eqnarray}
where
\begin{eqnarray}
\log {\Lambda (0) \over \Lambda (\theta)} = {5 \pi^2\over 11} 
{ 6 \tan \theta \over \left( 3 + 8 \tan \theta\right) }~~.~~
\label{lambda}
\end{eqnarray}
Here $\beta_0$ and $\beta_1$ are the usual first two universal coefficients 
of the $\beta$ function for the $SU(2)$ gauge theory:  they do not
depend on $\theta$.

One sees clearly from the equations above that the introduction of a
non-zero $\beta_A$, leads merely to a different $g_u$ and a
correspondingly different value for the scale $\Lambda(\theta)$.
However, each of these theories, including the usual Wilson theory for
$\beta_A = 0.0$ flow to the same critical fixed point, $g^c_u = 0$,
in the continuum limit and has the same scaling behavior near the
critical point.  The different forms of action, obtained by varying
$\beta_A$, are simply related by a redefinition of coupling constant and 
the intrinsic scale $\Lambda$ and yield the same universal continuum
physics.  Numerical investigations for different $\beta_A$ thus
constitute a necessary check of the finite cut-off effects in the
non-perturbative results obtained for $\beta_A=0.0$, i.e., the Wilson
action.

Bhanot and Creutz\cite{BhaCre} found that the lattice theory defined by 
the extended action has a rich phase structure (Fig. 1). Along the $\beta = 0$ 
axis it describes the $SO(3)$ model which has a first order phase transition 
at $\beta_A^{crit} \sim 2.5$.  At $\beta_A = \infty$ it describes the $Z_2$ 
lattice gauge theory again with a first order phase transition at 
$\beta^{crit} = {1\over2} \ell n(1 + \sqrt{2})$ $\approx$ 0.44 \cite {Weg}. 
Ref. \cite {BhaCre} found that these first order transitions extend into the 
($\beta$,$\beta_A$) plane, ending at an apparent 
critical point located at (1.5,0.9).  These transition lines are shown in 
Fig. 1 by continuous lines.  Using finite size scaling, Ref.
\cite{gavbulk} has recently shown that the critical endpoint must have
$\beta_A \ge 1.25$.  More simulations on larger lattices will be
required to determine the endpoint precisely.  The qualitative aspects of 
this phase diagram were also reproduced by mean field theory \cite {AlbFly} 
and large N\cite{OgiHor} and strong coupling\cite{DasHel} expansions.

Simulations of the mixed action above at finite temperature are made 
on asymmetric $N_\sigma^3 \times N_\tau$ lattices, with periodic boundary 
conditions in the (shorter) $\tau$-direction.  The partition function at
finite temperature is given by,
\begin{eqnarray}
Z = \int \prod_{x, \mu} dU_\mu(x)~~ \exp ( -S) ~~.~~
\label{parfun}
\end{eqnarray}
The order parameter for the deconfinement transition is the 
Polyakov loop \cite{McSv} defined by 
 \begin{eqnarray}
 L(\vec n) = {1\over2} Tr \prod^{N_\tau}_{\tau=1} U_0 (\vec n,\tau),
 \label{pol}
 \end{eqnarray}
Here $U_0 (\vec n,\tau)$ is the time-like link at the lattice site 
$(\vec n,\tau)$. 
Due to periodic boundary condition in the time-like direction at finite 
temperature the action of eq. (\ref{ea}) has a $Z_2$ invariance corresponding 
to the center of the gauge group. Defining this symmetry to be
 \begin{eqnarray}
 U_0 (\vec n,\tau_0) \rightarrow z U_0 (\vec n,\tau_0) ~~\forall n,~~ \tau_0 :
 {\rm fixed}~ {\rm ,~~and}~~ z ~~\in Z_2 ~~~~,
 \label{sym}
 \end{eqnarray}
one sees that under its transformation the Polyakov loop changes by 
 \begin{eqnarray}
 L \rightarrow z L~~~~ ,
 \label{tr}
 \end{eqnarray}
while the action in eq. (\ref{ea}) remains unchanged.

A non-vanishing value for $\langle L \rangle$, with respect to the
partition function in eq. (\ref{parfun}), signals a spontaneous
break-down of the global $Z_2$ symmetry.  $\langle L \rangle$ is also 
an order parameter for the deconfinement phase transition, as it 
(or equivalently its average value 
$L = {1 \over N_\sigma^3} \displaystyle \sum_{\vec n}
L(\vec n)$) can also be shown to be a measure of the free energy
of an isolated free quark \cite{McSv}. In order to monitor the 
critical behavior of the deconfinement transition, we also define 
the Polyakov loop susceptibility: 
\begin{eqnarray}
\chi_{N_\sigma} = N_\sigma^3 (\langle L^2 \rangle - \langle L \rangle^2)~~.~~
\label{chi}
\end{eqnarray}

In the thermodynamic limit, a second order transition is characterized 
by the following critical exponents: 
\begin{eqnarray}
\langle L\rangle & \propto & |T - T_c|^\beta  {\rm ~~~~~for}
			 ~~~~~{T \rightarrow T^+_c }\\[2mm]
\chi & \propto & |T - T_c|^{-\gamma}
	       {\rm ~~~for}~~~~~{T \rightarrow T_c }\\[2mm]
\xi & \propto &  |T - T_c|^{-\nu} {\rm ~~~for}~~~~~{T \rightarrow T_c}~~~~ .
\label{critexp}
\end{eqnarray}

Here $\xi$ is the correlation length corresponding to the Polyakov loop 
correlations and $\beta \approx 0.325$, $\gamma \approx 1.24$
and $\nu \approx 0.63$ are the Ising model exponents, assuming the
universality conjecture to be true.  The best determination of these
exponents for the $SU(2)$ lattice gauge theory was made\cite{EngFin} 
by using the finite size scaling theory\cite{Barb}, 
according to which, the peak of the $L$-susceptibility on a lattice
of spatial extent $N_\sigma$ is expected to grow like
\begin{equation}
\chi^{max}_{N_\sigma} \propto N_\sigma^{\omega}~~,~~
\label{chifs}
\end{equation}
where $\omega=\gamma/\nu=1.97$.
If the phase transition were to be of first order instead, then
one expects the exponent $\omega = 3$, corresponding to the dimensionality
of the space \cite{ChLaBi}.  In addition, of course, the order parameter is
expected to exhibit a sharp, or even discontinuous, jump and the corresponding
probability distribution should show a double (multi) peak structure.
For $\beta_A = 0$, the universality prediction was verified
by Monte Carlo simulation
by Engels et. al. \cite {EngFin}, who found $\omega= 1.93 \pm 0.03$,
whereas we found\cite{us2} $\omega = 3.25 \pm 0.24$ for $\beta_A =
1.4$ on $N_\tau = 2$ lattices.

\bigskip 

\begin{center}
3.  \bf STRONG COUPLING \\
\end{center}
\bigskip

Before turning to the results of our simulations to determine $\omega$
and to locate the tricritical point, it may be an instructive exercise
to find out what hints the strong coupling expansion can provide.
Such expansions for the free energy \cite{PolSzl,GreKar} and string tension 
\cite{Gre} have been used in the past to study $SU(N)$ deconfinement 
transition for the Wilson action.  The basic strategy is to obtain an
effective potential for the order parameter $L$, by expanding the
partition function in powers of the inverse coupling constant(s) and
integrating out the spatial links .  Due to the $Z_2$
symmetry of the theory, the Landau-Ginzburg effective action is an even 
polynomial in the Polyakov loop for the $SU(2)$ theory.
To lowest order:  
	    
\begin{eqnarray}
S_{eff} & = & -{1 \over 2} \sum_{\vec{n}} 
	   \log\left(1-L^2\left(\vec{n}\right)\right) 
	    -  4\left({\beta \over 4}\right)^{N_\tau} \sum_{\vec{n},i} 
	    L\left(\vec{n}\right)L\left(\vec{n}+i\right) \nonumber \\ [2mm]
	& - & \left({\beta_A \over 9}\right)^{N_\tau} \sum_{\vec{n},i} 
	    \left(4 L^2\left(\vec{n}\right) -1\right)
	    \left(4 L^2\left(\vec{n}+i\right)-1\right)   
\label{lg}
\end{eqnarray}

Here the first term is independent of the couplings and is the exact Jacobian 
due to the change of the temporal link variables to $L(\vec{n})$ after all 
the  link 
integrations. The last two terms are the leading strong coupling terms in 
$({\beta \over 2})$ and $({\beta_A \over 3})$ with the  assumption that both 
$\beta \over 2$ and $\beta_A \over 3$ are small and treated on 
the same footing. Only the leading order terms in $\beta$ and $\beta_A$ 
are retained in the effective action here.  Demanding translational
invariance for the configuration which minimizes the action, one can
easily obtain the effective potential for the order parameter.
Expanding the log term for small $L$, one has the following results
for the coefficients, $b_2$, $b_4$ and $b_6$ of the $L^2$, $L^4$ and
$L^6$ terms in the effective potential:
\begin{eqnarray}
 V_{eff}(L) = b_2 L^2 + b_4 L^4 + b_6 L^6~~,~~
\label{veff}
\end{eqnarray} 
where
\begin{eqnarray}
 b_2 & = & {1 \over 2} - 12 \left({\beta \over 4}\right)^{N_\tau} + 
  24 \left({\beta_A \over 9}\right)^{N_\tau} \\  
 b_4 & = & {1 \over 4} - 48 \left({\beta_A \over 9}\right)^{N_\tau} \\
 b_6 & = & {1 \over 6}~~.~~
\label{tc}
\end{eqnarray} 

The positivity of the $b_6$-term (and all other higher terms) ensures
that the effective potential is bounded from below. Note that for small
enough $\beta$ and $\beta_A$, $b_2$ and $b_4$ are also positive,
favoring thus the confined phase of $L=0$. For small $\beta_A$ and
arbitrary $\beta$, $b_4$ remains positive but $b_2$ goes
through a zero, giving rise to a second order phase transition at a
critical $\beta$ obtained by setting $b_2$ to zero.  As $\beta_A$
increases, $b_4$ becomes negative above a critical value 
of $\beta_A$.  The effective potential then has two additional minima 
in addition to the one at $L=0$. As $\beta$ increases, these minima
deepen and become equal to the one at $L=0$, yielding a first order 
deconfinement phase transition.  The  tricritical point, where the
deconfinement phase transition changes to become first order, is given by  
setting the coefficients $b_2$ and $b_4$ to zero.

In this leading order strong coupling expansion, the tricritical points are 
$(\beta^{tricrit}, \beta_A^{tricrit})$ = (0.913,0.649), (1.91,2.418) for 
$N_\tau$=2 and 4 lattices respectively. This suggests  that the tricritical 
point moves towards the top right corner $(\beta=\infty,\beta_A=\infty)$ of 
the phase diagram as the lattice spacing is reduced.   Thus 
the the tricritical point will not seen by the continuum limit. 
Of course, one needs to improve the leading order strong coupling result
as $N_\tau$ increases and check that this conclusion remains unchanged.
Nevertheless, these results are encouraging for two reasons.  Firstly,
they provide a concrete example of how the $SU(2)$ gauge theory at
finite temperature $can$ have a first order phase transition.  Indeed,
it should be noted that the bulk phase transition plays no role above in
changing the order of the deconfinement phase transition at large
$\beta_A$.  Secondly, the qualitative trend suggested by this simple
exercise is in agreement with the naive idea of independence of physical
results with respect to irrelevant couplings.  Of course, the key
question of the limit of the coefficients $b_i$, as the lattice spacing
$a \rightarrow 0$, can only be resolved by simulations at present and, in
principle, the trajectory of the tricritical point could go either way
in that limit.  In the next section, we describe the results of our
simulations which were made in an attempt to answer this issue.

\bigskip
\begin{center}
4.  \bf RESULTS OF THE SIMULATIONS\\
\end{center}
\bigskip
 
Our Monte Carlo simulations were done using Metropolis algorithm
on $N_\sigma^3 \times N_\tau$ lattices with $N_\tau$=2, 4, 6, 8 and 
$N_\sigma$= 8,10,12,16. The many different values of 
$N_\sigma$ were chosen to study the finite size scaling behavior of the 
theory and to compute the critical exponent $\omega$, while the $N_\tau$
values were chosen to monitor the movement of the tricritical point
with decreasing lattice spacing.  We also used $N_\sigma$= 16 in one case 
to be sure of the critical exponent.  The possible ranges for the tricritical 
points for different $N_\tau$ were known from our earlier work, and the 
simulations were carried out at $\beta$ and $\beta_A$ in these ranges.  
Histogramming techniques were used to extrapolate to nearby $\beta$ values 
while estimating the height and location of the peak of various
susceptibilities.  The values of the critical exponents from our earlier 
simulations\cite{us1,us2} on $N_\tau$=2, 4 lattice are summarized in Table 1,
where the result of Ref. \cite{EngFin} for $\beta_A=0.0$ is also given.
One sees a good agreement with the Ising model exponent for $ N_\tau=4$
for $\beta_A \le 0.9$, suggesting that the tricritical point must lie at
higher $\beta_A$ in this case.  On the other hand,
no check of the universality with the three dimensional Ising model has so 
far been made for the $N_\tau$=2 lattices:  the corresponding $\omega$
is unknown although all earlier simulations do indicate a continuous 
transition.  From our previous work\cite{us2}, we know that the transition for
$N_\tau=2$ and for $\beta_A \ge 1.4$ is a strong first order one.  For 
$\beta_A=1.4$, the exponent $\omega$ = 3.25(24), with a typical  tunneling 
time of $\approx 30,000-40,000$ Monte Carlo sweeps. 
Thus the effective ranges for the $\beta_A$ of the tricritical points were
$ 1.4 > \beta_A^{tricrit}$ for $N_\tau = 2$, and $\beta_A^{tricrit} >
0.9$ for $N_\tau=4$.  No firm upper bound was known for the 
latter case, although we had good indications that $1.5 >
\beta_A^{tricrit}$, as Ref. \cite{us1} found a
co-existing two state signal in both the Polyakov loop, $L$, and the
plaquette, $P$, at $\beta_A = 1.5$.

In our earlier work, the simulations at $\beta_A$=1.1 on the $N_\tau=2$
lattices did not reveal a clear three peak structure in the histogram of the 
Polyakov loop on the $N_\sigma =8,10,12$ lattices,
although the peaks did become a bit sharper on going to the $N_\sigma=12$ 
lattice.  Correspondingly the determination of the critical 
exponent did not fix the order of the transition uniquely.  This is similar 
to  the $N_\tau$=4 results at $\beta_A$=1.1 \cite{us1} where the histograms 
and the evolution graphs of the Polyakov loop gave a very weak two state 
signal with an $\omega \simeq 2.34$, lying between values
characteristic of first and second order phase transition.
Such a behavior of the deconfinement transition 
can be understood from the point of view of the effective potential in 
terms of the Polyakov loop, if these simulations were indeed close to
the tricritical point.  As argued in earlier section, the first two 
leading coefficients of $L^2$ and $L^4$ terms of the effective potential 
are then close to being zero, leading to a reasonably flat effective
potential around $L$=0.0, a fact which we will later exploit to conclude 
about the movement of the tricritical point as the lattice spacing is reduced. 
However, as a consequence, much larger statistics is required to sample 
the exact nature of the effective potential near the tricritical point  
to separate a weak first order transition from a second order one. 
We therefore increased the statistics to typically $4 \times 10^6$ sweeps 
to compute the critical exponents and focused more on $\beta_A$ close to 1.1.
We, however, also made simulations on $N_\tau$=2 lattices at $\beta_A$=0.0,
0.8, and 0.9 to determine the critical exponent $\omega$ and thus the
range for the tricritical point more precisely.
The observables were typically recorded after every 20 sweeps to 
reduce auto-correlation. The errors were estimated by further binning
the data and the typical bin size was $O(100)$.

Figs. 2a, 2b, 2c and 2d exhibit the results for $L$-susceptibility for
the $N_\tau=2$ lattices for $\beta_A=0.0$, 0.8, 1.1 and 1.25 respectively. 
The results for $\beta_A=0.9$ are similar to those in Fig. 2c and are therefore
not shown here. In each of these figures, the range of expected values for the 
peak height for the $N_\sigma$=10 and 12 lattices is also shown by two 
horizontal lines by assuming the validity of eq. (\ref{chifs}), $\omega=1.97$, 
and by using $\chi^{max}_{N_\sigma=8}$ for each case.  The errors on the 
respective $\chi^{max}_{N_\sigma=8}$ induce the spread between the
lines.  We always chose a fresh starting point in an iterative manner,
if the initial guess was too far away from the extrapolated estimate
for the location of the peak.  This reduced the influence of
the unknown systematic errors in the peak height due to our extrapolation 
procedure.  As seen in Figs. 2a-2d, we do hope that this 
source of errors has been brought under control by our choice of
the simulation points, and that it does not annul our conclusions about 
whether the critical exponent is close to 3 or 1.97.

The independence of $\omega$ for $\beta_A =0.0$ (and small $\beta_A$)
for $N_\tau$ =2 and 4 and its agreement with 3-d Ising model value is
satisfying since there is only one known critical point in $Z(2)$
symmetric theories and $a~priori$ one expects the exponents to be universal.  
Moreover, strong coupling arguments, which predict a second order phase 
transition for small $\beta_A$, should be more reliable for smaller $N_\tau$. 
Quantitatively, however, one notices the leading order strong coupling 
prediction for  
$\beta^{crit}(N_\tau=2)$= 0.816 for $\beta_A=0.0$ to be far away from the 
corresponding Monte Carlo determination.  Furthermore, its $\beta_A$-dependence
seems to be also in the wrong direction.  Thus, one really could have
expected surprises in form of a qualitative difference from the strong 
coupling prediction as well.  On the other hand, it may be more natural
to expect the effect of higher orders in quantitative shifts and not in
qualitative features.  Since it is unclear whether $N_\tau$=4 is in the
strong coupling region, the universality conjecture for 
the critical exponents needs to be tested on lattices with larger
$N_\tau$ and thus closer to the continuum limit even for the Wilson
action, i.e, $\beta_A = 0.0$.

The values of the critical coupling, $\beta^{crit.}$  and the finite size 
scaling exponent, $\omega$, obtained by fitting the peak heights to 
eq.(\ref{chifs}), are given  in the Table 2 for all the $\beta_A$ values we
investigated, including $\beta_A=0.9$. These estimates of $\omega$ in Table 2, 
along with the agreement in Fig. 2 with the predictions based on a 
$\omega=1.97$ show the deconfinement phase
transition for  $N_\tau = 2$ lattices to be a clear second order with
Ising model exponents for  $1.25 \ge \beta_A \ge 0.0$ .  
At $\beta_A$=1.25, additional simulations were performed on $N_\sigma$=16 
to confirm the second order nature of the transition. Note that if
the predictions for $\omega$=3, corresponding to a first order phase
transition were to be displayed in Figs. 2 then they would overshoot by
a large amount, especially in Fig. 2d where they would be too big by a
factor of 2 for the $N_\sigma=16$ lattice.  Also interesting to note is
the resolution of the ambiguity in establishing the order of the phase
transition for $\beta_A=1.1$ in these better statistics simulations.
The  finite size scaling exponent is $\omega$= 1.79(02) and is thus a lot
closer to the Ising value.  The tricritical point T on a $N_\tau$=2 lattice 
is thus constrained to lie in the range $1.4 \ge \beta_A > 1.25$, as
indicated in Fig. 1 by the gap between the filled circles for the first 
order transition points and hollow circles for the second order transition 
points.  Thicker dashed and dotted lines show the first order and second order
deconfinement phase transition lines.

Since the exponent $\omega$ was found\cite{us1}to be 2.31(28) for 
$N_\tau = 4$ at $\beta_A=1.1$ and the histograms of the Polyakov loop 
signaled a very weak first order transition,  we chose to re-investigate the 
transition first at $\beta_A=1.1$ and then move to larger values of $\beta_A$.  
Fig. 3 shows the evolution of both $L$ and $P$ at 
$\beta_A$ =1.25 on $N_\sigma$=8, 10, 12 lattices. These figures clearly show 
the coexistence of two states at the deconfinement transition, since
$\langle L \rangle \simeq 0$ for one of the phases while it is nonzero
and large for the other.  The plaquette, P, has a discontinuity at the
same location, and further the number of tunnelings and duration in
each phase do indeed decrease as the spatial volume increases.
Figs. 4a and 4b show the Polyakov loop susceptibility at $\beta_A$=1.1 and 
1.25 on $N_\tau$ = 4 lattice for various spatial volumes.  As in Figs.
2, the expected peak heights for the bigger two lattices are shown by
horizontal bars.  The only difference here is that the solid  horizontal
lines in Fig. 4-b show the predicted $\chi^{max}_{N_\sigma=10,12}$ 
by assuming $\omega$ = 3.0, while the broken lines in Fig. 4b and the
solid lines in Fig. 4a are for the Ising value 1.97.
The values of  couplings at which simulations were performed along 
with the critical values and the fitted $\omega$ are given in Table 3. 
The most astonishing result is that the transition at $\beta_A$ 
=1.25 is  a first order transition with $\omega= 3.13(01)$.
This needs to be contrasted with i) the results for $N_\tau=2$ and in
particular, the Fig. 2d and ii) the results of Ref. \cite{gavbulk} where
a clear absence of a first order bulk phase transition at $\beta_A$
=1.25 was shown.

These simulations thus indicate that the tricritical
point for $N_\tau$= 4 lies definitely {\it below} $\beta_A=1.25$ whereas
the corresponding $N_\tau=2$ point is definitely {\it above} $\beta_A=1.25$.
This is also clearly seen in Fig. 1, where again the first order transition 
points for $N_\tau=4$ are shown by filled squares, the second order
points by hollow squares and the gap between them is the allowed range for the
tricritical point.  As one can see, the tricritical point does shift as
the temporal lattice size increases from 2 to 4.  However, the direction
of the shift is almost orthogonal to the strong coupling prediction of the
previous section and its magnitude is also much smaller.
The shift, on the other hand, suggests a possible lack of any
correlation of the bulk transitions, if any, with the key observation of
the change of the order of the deconfinement phase transition.  This is
so since any possible bulk transition for $\beta_A \le $ 1.25 is definitely 
not\cite{gavbulk} a first order phase transition, making it an unlikely 
cause of the behavior seen in Figs. 3 and 4.  A plausible explanation 
then is indeed the phenomenon seen in the strong coupling calculations
in Sec. 3, i.e., a change of sign in the coefficient $b_4$ of the
effective potential for the order parameter, $L$.

The above downward movement of T in the extended coupling plane is 
puzzling. It  calls for a more detailed cross-check on the 3-d Ising model 
universality, especially for positive values of $\beta_A$ but also perhaps 
for larger $N_\tau$ at $\beta_A=0.0$.  Note that the latter has so far been
demonstrated to a very good accuracy only on the $N_\tau=2,4$ lattices. 
Therefore, we decided to monitor the deconfinement transition further by 
simulating the model on $N_\tau=6$ and $N_\tau$=8 lattices with 
$N_\sigma$=12 and 16 respectively.  On these lattices the simulations were 
performed at $\beta_A$=1.1 and 1.25 to determine the range in which T may
lie. At $\beta_A$=1.25 on $N_\tau$ =6 lattice we found the transition to 
be first order. The corresponding histogram is plotted in Fig. 5 at 
$\beta$=1.2184. The three peaks are clearly visible and distinct, though 
not of equal height.  This figure suggests the transition point to be at
a slightly higher value of $\beta$ than 1.2184.  In choosing this
$\beta$ we were guided by the location of the peak of the 
$L$-susceptibility to locate the critical point. We have found that this 
criterion to determine the critical point differed a little from that of 
the effective potential picture (i.e, the nature of the $L$-histograms), 
although they will coincide in the thermodynamic limit. 
Comparing the shapes of the 
histograms at $\beta_A$=1.1 in Figs. 6-a,b,c ($N_\tau$=4,6,8 respectively)  
on the other hand, one observes that their profiles at the peak 
tend towards a Gaussian behavior as one makes the 
lattice spacing smaller by going from $N_\tau=4$ to 6 to 8. 
This is the expected behavior if the tricritical point shifts upwards from $\beta_A 
\simeq 1.1$ for $N_\tau=4$. Since the effective potential (\ref{lg}) 
will have $b_2 \approx b_4 \approx 0$ at the tricritical point, it will
have a reasonably flat bottom, causing the very flat top for the $N_\tau$=4 
histogram. In Fig. 6-a we have plotted the histograms for $N_\tau$=4 and 
$N_\sigma$=8,10, 12 to confirm the closeness of the tricritical 
point with the run point, i.e. $\beta$=1.327 and 1.32685. Moreover, one 
clearly sees that the flatness is not a finite $N_\sigma$ artifact. The 
reduction in the above flatness would signal $b_4$
becoming nonzero, as in a typical second order transition.  More
quantitatively, the fluctuations in the Polyakov loop at 
$\beta_c= 1.327$ on the $N_\tau$ =4 lattice, are in the 
range $\Delta L \approx \pm$ 0.3, while $\Delta L \approx \pm$ 0.1
on the $N_\tau=6$ lattice at $\beta_c$=1.339, with a reduction by a
factor of 3 in the flatness of the histogram at the top. The statistics 
for the above runs on $N_\tau$ =4 and 6 lattices was roughly $2 \times 10^6$
and $7 \times 10^6$ respectively.  In principle, 
this could be taken as a hint that the tricritical point has moved up.   
However, one knows that $L$ measures the free energy of a point-like 
test charge and thus has $N_\tau$-dependent corrections.  These reduce 
the value of $L$ just above the phase transition as $N_\tau$ increases.  A
perturbative estimate of these corrections will presumably still not yield 
the factor of two but unless it is quantitatively so demonstrated
one is handicapped in drawing a firm conclusion about lack of the finite 
$N_\tau$ effects in the flatness of the effective potential.
The Polyakov loop histogram in Fig. 6-c corresponding to 
$N_\tau=8$ has no indication of the co-existence of 2 phases and therefore 
no evidence of a tricritical point in its vicinity.  In this figure we have 
plotted 2 histograms corresponding to $\beta$=1.35 and 1.3508 showing the 
the sharp behavior of the transition. This  indicates that the qualitative 
behavior of the histogram changes drastically with the coupling. This  again
is not expected near the tricritical region.  The above drastic reduction in 
the flat region compared to $N_\tau$ =4 lattice could again be partially due to 
the same cause as above.   Nevertheless, these
qualitative observations suggest that that the tricritical point T
moves upwards as $N_\tau$ is increased from 4 to 6 and then from 6 to 8. 
Even assuming that this upward motion of the the tricritical 
point continues, one will need a lot larger lattices to confirm 
universality for $\beta_A > 0$ since the upward movement is rather small.  

\bigskip

\begin{center}
5.  \bf SUMMARY AND DISCUSSION \\
\end{center}
\bigskip

We simulated the extended action of eq.(\ref{ea}) on $N_\tau=2$, 4, 6 
and 8 lattices with varying spatial sizes and determined the order of the
deconfinement phase transition by obtaining the critical exponent $\omega$ 
using finite size scaling theory and also by observing the behavior of the 
histograms.  For $\beta_A=0.0$ and $N_\tau=2$, we find $\omega=1.92 \pm 0.01$ 
which is in good agreement with the corresponding Ising model
value $1.97 \pm 0.03$. Thus the deconfinement phase transition for the
Wilson action for both $N_\tau=2$ and 4 is of second order and is in the
same universality class as the three-dimensional Ising model.  Our
simulations show this universality to persist when $\beta_A$ is turned
on.  For the $N_\tau=2$ lattice, this is true for at least $ \beta_A \le
1.25$ while for $N_\tau=4$ the phase transition at $\beta_A=1.25$
is already of first order, with $\omega = 3.13 \pm 0.01$.  The tricritical 
point where the deconfinement phase transition changes it order is therefore 
definitely above $\beta_A = 1.25$ for $N_\tau=2$ while it is definitely below
$\beta_A=1.25$ for $N_\tau=4$ lattices.  Placing it in the middle of the
ranges we have narrowed down, it shifts vertically downwards by about
0.15.  There is a horizontal positive shift in $\beta$, of the order of 0.09, 
as well.  

The comparison of the above shift of  the tricritical point on $N_\tau$=2 
and 4 lattices  with the leading order strong coupling prediction shows
surprises.  It is, of course, reassuring that the change of the
order and the existence of tricritical points for each $N_\tau$ is
as predicted.  However, the predictions do very poorly on a quantitative
level.  In particular, the directions of the shift of the phase
transition as $\beta_A$ is turned on, and more importantly, the
predicted vertically upward shift of the critical point are in complete
contrast with the simulation results for $N_\tau$ =2 and 4.  On the other hand,  
the leading order strong coupling equations are known to fail 
quantitatively for the Wilson theory for both
$N_\tau=2$ and 4 as well.   Putting $\beta_A$=0.0 in 
eq. (\ref{tc}), we recover \cite{PolSzl}  the criticality condition for the 
Wilson action : $\beta^{critical} = 4 (1/24)^{1 \over N_\tau}$. 
Therefore, the values of the critical couplings 
to this order on $N_\tau$ = 2 and 4 lattices are 0.816 and 1.807
respectively.  These should be compared with the corresponding Monte
Carlo values of this work ($\beta_c(N_\tau=2) = 1.88(01)$) and of Ref. 
\cite{EngFin}($\beta_c(N_\tau=4) = 2.30(01)$).  It would be interesting
to check whether the inclusion of higher orders in the strong coupling
expansion can yield a better agreement with the simulation results, 
especially for the direction of the vertical shift of the tricritical
point.

The above downward movement of the tricritical point on going from $N_\tau$=2 
to 4 was also observed in Ref. \cite{Ste} where the Villain form of eq. 
(\ref{ea}) was simulated.  
This unexpected behavior of the tricritical point may therefore need to
be taken seriously. 
If this trend persists even on larger lattices then 
the continuum limit of SU(2) lattice gauge theory could correspond to 
a first order deconfinement transition. Hence we simulated this model on 
larger lattices with $N_\tau$ =6 and 8 and found some hints of an upward 
movement of the tricritical point by comparing the shapes of the Polyakov 
loop histograms.  While it unfortunately is not very conclusive,
it is encouraging that the trajectory of the 
tricritical point could potentially be turning up in the coupling plane
for these $N_\tau$.  The tricritical point for both 
these lattices was still found to be below $\beta_A$=1.25 
and above $\beta_A = 1.1$.  Much more computational resources on 
bigger lattices with $N_\tau $ much bigger than 8 are necessary to
confirm this.

It may therefore be important to understand and 
explain the origin of the change in the order of the deconfinement 
transition even away from the continuum, especially since 
the results of Ref. \cite{gavbulk} suggest a lack of a  first
order bulk phase transition at $\beta_A=1.25$ for this action and thus
make it implausible that a bulk transition is responsible for such a change.
To answer this question, we plan to consider 
the Villain form for the SO(3) part of the extended action\cite{CanHalSch}:  
 
\begin{eqnarray}
Z = \sum_{\sigma_p(n)=\pm 1} \int \prod_{\mu, n} dU_\mu(n)\exp [
\sum_p\left({\beta \over 2} + {\beta_A \over 3} \sigma_p\right) Tr_F U_p +
\lambda \sigma_p ]~~,~~
\label{vil}
\end{eqnarray}

\noindent
Here  $\sigma_p$ is a $Z_2$ plaquette field and the summation over it 
ensures the invariance of the second term above under $U_\mu(n) \rightarrow
 - U_\mu(n)$. For $\lambda=0$, this action is again in the same universality 
class as that of (\ref{ea}). In fact as already mentioned, its 
simulations\cite{Ste} 
on  $N_\tau=2,4$ lattices led to exactly the same behavior of the tricritical 
point as reported in this paper. Besides computational advantages \cite{Ste}, 
the theoretical advantage 
of this action is that unlike eq.(\ref{ea}), the $SO(3)$ monopoles and their 
dynamics is manifest in the form of the $Z_2$ plaquette field. 
The SO(3) monopole charge density is given by $\prod_{p \in c}\sigma_p$;  
here the product is over the 6 faces of a cube \cite{Tom2}.   
 The last term in the above 
equation is the potential energy for these topological degrees of freedom. 
In the extreme ($\lambda \rightarrow \infty$) case when all the $SO(3)$ 
monopoles are suppressed ($ \forall \sigma_p = +1$), 
the above extended action reduces to Wilson action with redefined coupling 
and therefore has only a second order deconfining transition on small
$N_\tau$ lattices. Therefore, these topological degrees of freedom may
have a crucial role in changing the order of the transition. 
In the extended coupling plane these
monopoles cost less and less energy as the adjoint coupling is increased
with decreasing values of $\beta$. Therefore, above $\beta_A^{tricritical}$ 
they might condense giving rise to a first order transition. This can be 
checked by simulating the above model.  If true, this phenomenon will be 
particularly appealing because precisely the same degrees of freedom and 
their condensation have been attributed to the first order nature of the 
`bulk transition' which we find to be first order deconfinement transition.  
It may thus also enable in resolving the physical nature of the transition.

\bigskip 

\begin{center}
6.  \bf ACKNOWLEDGMENTS \\
\end{center}
\bigskip 

The computations reported here were performed on the DEC Alpha
machines of the Tata Institute of Fundamental Research, Bombay and the
Theoretical Physics Institute, University of Minnesota, Minneapolis. We
would like to thank the staff at these institutes for their support. One
of us (R.V.G.) gratefully acknowledges the hospitality he received in
TPI, Minneapolis, especially from Profs. J. Kapusta and L. McLerran.
This work was supported by the U. S. Department of Energy under the
grant DE-FG02-87ER40328. M.M acknowledges the I.N.F.N. fellowship. 
We wish to thank Prof. M. Grady, SUNY, Fredonia, for pointing out an 
error in the earlier version of the manuscript.

\newpage 
\pagestyle{empty}
\begin{table}
\caption[]{
The average values of the critical exponent $\omega$ at different values of the adjoint 
couplings.  The expected value  is 1.97 (3.0) if the deconfining phase transition is second 
order (first order). The data taken from Refs. \cite{EngFin,us1,us2}
}

\vspace{1.0cm}
\begin{tabular}{ccc}
$\beta_A$ & $N_\tau$ & $\omega$ \\ \hline
0.0              &    4      &      1.93(03)     \\ \hline
0.5              &    4      &      1.92(29)     \\ \hline
0.75             &    4      &      1.53(32      \\ \hline
0.9              &    4      &      2.10(22)     \\  \hline 
1.1              &    4      &      2.34(15)     \\   \hline
1.4              &    2      &      3.246(243)   \\  
\end{tabular} 
\label{table1} 
\end{table}

\pagestyle{empty}

\begin{table}
\caption[]{
The values of ($\beta,\beta_A$) on $N_\sigma^3 \times 2$ lattice 
at which simulations were 
performed, $\beta^{crit.}$ and the finite size scaling exponent $\omega$.  The expected 
value for $\omega$ is 1.97 (3.0) if  the deconfining phase transition is second order (first order). 
}

\vspace{1.0cm}
\begin{tabular}{ccccc}
$\beta_A$ & $N_\sigma$ & $\beta$ & $\beta_c$ & $\omega$ \\ \hline
    & 8  & 1.90  & 1.88  &           \\
0.0 & 10 & 1.885 & 1.878 & 1.92(01)  \\
    & 12 & 1.877 & 1.877 &          \\
\hline
     & 8   & 1.368  &1.368  &         \\  
0.8  & 10  & 1.367  &1.3664 &  2.03(01) \\
     & 12  & 1.368  &1.366  &           \\
\hline
     & 8  & 1.3    &  1.31    &          \\  
0.9  & 10 & 1.3    &  1.31    & 1.83(02) \\  
     & 12 & 1.3092 &  1.3088  &          \\  
\hline
       &   8   & 1.201  &   1.201    &          \\
1.1    &   10  & 1.2    &   1.2      & 1.79(02) \\
       &   12  & 1.1999 &   1.1995   &          \\
\hline
     &   8   &  1.12    &1.1203    &          \\
1.25 &  10   &  1.12    &1.1203    &          \\
     &  12   &  1.12    &1.12      & 1.97(01) \\
     &  16   &  1.12    &1.1196    &          \\
\end{tabular} 
\label{table2} 
\end{table}
\newpage

\newpage 
\pagestyle{empty}

\begin{table}
\caption{
The values of ($\beta,\beta_A$) on $N_\sigma^3 \times 4$ lattice at which simulations were 
performed, $\beta^{crit.}$ and the finite size scaling exponent $\omega$.  
}

\vspace{1.0cm}
\begin{tabular}{ccccc}
$\beta_A$ & $N_\sigma$ & $\beta$ & $\beta_c$ & $\omega$ \\ \hline
    & 8  & 1.327   & 1.327   &          \\ 
1.1 & 10 & 1.327   & 1.3274  & 2.11(02) \\
    & 12 & 1.32715 & 1.32715 &          \\ 
\hline
     & 8   & 1.2146  & 1.2142 &           \\ 
1.25 & 10  & 1.214   & 1.2144  & 3.13(01) \\
     & 12  & 1.2144   & 1.2143  &         \\  
\end{tabular} 
\label{table3} 
\end{table}

\newpage
\pagestyle{empty}
\begin{center}
\bf{FIGURE CAPTIONS}
\end{center}

\bigskip

\noindent Fig. 1 
The phase diagram of the extended SU(2) lattice gauge theory.  
The solid lines are from simulations done on a $5^4$ lattice by 
Bhanot and Creutz\cite{BhaCre}.  The light dashed line indicates 
the absence of first order bulk transition \cite{gavbulk} below 
$\beta_A=1.25$.  The dotted(thick dashed) lines with 
hollow(filled) symbols are the second(first)  order deconfinement 
phase transition lines on $N_\tau$ =2 (circles) and 4 (squares) lattices.  
\bigskip

\noindent Fig. 2
 Polyakov loop susceptibility at (a) $\beta_A$=0.0, (b) 
$\beta_A$=0.8, (c) $\beta_A$=1.1 and (d) $\beta_A$=1.25  on $8^3 \times 2$, 
 $10^3 \times 2$ and $12^3 \times 2$ lattices.  At $\beta_A=1.25$  result 
 on $16^3 \times 2$ lattice is also shown. The points with error bars are 
 results of simulations and the curves are extrapolations by the 
 histogramming technique. The horizontal lines are predictions 
assuming a second order deconfinement transition, as explained in the text. 
\bigskip

\noindent Fig. 3
Evolution of $|L|$ and $P$ at $\beta_A = 1.25$ on 
(a) $12^3 \times 4$ ($\beta =1.2144$), (b)$10^3 \times 4$ 
($\beta=1.214$) and (c)$8^3 \times 4$ ($\beta=1.2146$) 
lattices.
\bigskip

\noindent Fig. 4
Polyakov loop susceptibility at (a)$\beta_A$=1.1 and 
(b)$\beta_A$=1.25 on $8^3 \times 4$, $10^3 \times 4$ and $12^3 
\times 4$ lattices.  The solid horizontal lines are predictions 
assuming a first (second) order deconfinement transition at 
$\beta_A$ = 1.25 ($\beta_A$=1.1), as explained in the text. 
The broken lines in Fig. 4a are predictions for second order.
\bigskip

\noindent Fig. 5 

The probability density of L at $\beta_A=1.25$ on $12^3 \times 6$ lattice at $\beta$=1.2184.  

\noindent Fig. 6

The probability  density of L at $\beta_A=1.1$ on $N_\sigma^3 \times N_\tau$ lattice for 
a] $N_\tau$=4, $N_\sigma$=8,10,12 and $\beta$=1.327,1.327 and 1.32685 respectively, 
b] $N_\tau$=6, $N_\sigma=12$ and $\beta$= 1.339, c] $N_\tau$=8, $N_\sigma=16$ and 
$\beta$= 1.35 and 1.3508.


\newpage


\begin{references}
\bibitem{Wil} K. Wilson, Phys. Rev. {\bf D10} (1974) 2445.
\bibitem{HalSch} G. Halliday, A. Schwimmer, Phys. Lett. {\bf 101B} (1981) 327 
and B. Lautrup, Phys. Rev. lett {\bf 47} (1981) 9, M. Creutz, Phys. Rev. Lett. 
{\bf 46} (1981) 1441. 
\bibitem{BhaCre} G. Bhanot and M. Creutz, Phys. Rev. {\bf D24} (1981) 3212.
\bibitem{CanHalSch} L. Caneschi, I. G. Halliday and A. Schwimmer, Nucl. Phys.
 {\bf B200 [FS4]} (1982) 409.
\bibitem{AlbFly} J. M. Alberty, H. Flyvbjerg and B. Lautrup, Nucl. Phys. 
{\bf B220[FS8]} (1983) 61.
\bibitem{OgiHor} M. C. Ogilvie, A. Horowitz, Nucl. Phys. {\bf B215[FS7]} (1983) 249.
\bibitem{DasHel} R. Dashen, Urs M. Heller and H. Neuberger, Nucl. Phys.
{\bf B215[FS7]} (1983) 360.
\bibitem{Tom1} E. T. Tomboulis Phys. Lett.  {\bf 303B} (1993) 103.
\bibitem{Tom2} E. T. Tomboulis Phys. Lett.  {\bf 108B} (1982) 209.
\bibitem{BroKes} R. C. Brower, D. A. Kessler and H. Levine, Nucl. Phys.
{\bf B205[FS5]} (1982) 77.
\bibitem{FinHelKar} J. Fingberg, U. Heller and F. Karsch, Nucl. Phys. 
{\bf B392} (1993) 493.
\bibitem{EngFin} J. Engels, J. Fingberg and M. Weber, Nucl. Phys. 
{\bf B332} (1990) 737; \\
J. Engels, J. Fingberg and D. E. Miller, Nucl. Phys. 
{\bf B387} (1992) 501.
\bibitem{SveYaf} B. Svetitsky and L. G. Yaffe, Nucl. Phys. 
{\bf B210[FS6]} (1982) 423.
\bibitem{us1} R. V. Gavai, M. Grady and M. Mathur Nucl. Phys.
{\bf B423} (1994) 123.
\bibitem{us2} M. Mathur, R. V. Gavai, Nucl. Phys.  {\bf B448} (1995) 399;
Nucl. Phys. B (PS) 42 (1995) 490. 
\bibitem{Ste} P. Stephenson, hep-lat/9509070.
\bibitem{EngSch} J. Engels and T. Scheideler, hep-lat/9610019. 
\bibitem{BhaDas} G. Bhanot and R. Dashen, Phys. Lett.
{\bf 113B} (1982) 299.
\bibitem{MuSch} K.-H. M\"utter and K. Schilling, Phys. Lett.
{\bf 121B} (1983) 267.
\bibitem{Gav} R.V. Gavai, Nucl. Phys. {\bf B215 [FS7]} (1983) 458.
\bibitem{GavKarSat} R.V. Gavai, F. Karsch and H. Satz, Nucl. Phys. {\bf B220 
[FS8]} (1983) 223.
\bibitem{gavbulk} R.V. Gavai, Nucl. Phys. {\bf B474 } (1996) 446
\bibitem{KorAlt} A. Gonzalez-Arroyo and C. P. Korthals-Altes, Nucl. Phys. 
{\bf B205} (1982) 46.
\bibitem{Weg} F. J. Wegner , J. Math. Phys. {\bf 12} (1971) 2259.
\bibitem{McSv} L. McLerran and B. Svetitsky, Phys. Rev. {\bf D24}
 (1981) 450.
\bibitem{Barb} M. N. Barber, in Phase Transitions and Critical Phenomena, 
 vol. 8, Ed. C. Domb and J. L. Lebowitz (Academic Press, New York, 
1983) p. 146.
\bibitem{ChLaBi} M. S. Challa, D. P. Landau and K. Binder, 
Phys. Rev. {\bf B34} (1986) 1841.
\bibitem{PolSzl} J. Polonyi, K. Szlachanyi, Phys. Lett. {\bf 110B} (1982) 395.
\bibitem{GreKar} F. Green, F. Karsch Nucl. Phys. {\bf B238}  (1984) 297.
\bibitem{Gre} F. Green, Nucl. Phys. {\bf B215 [FS 7]}  (1983) 83.
\end{references}
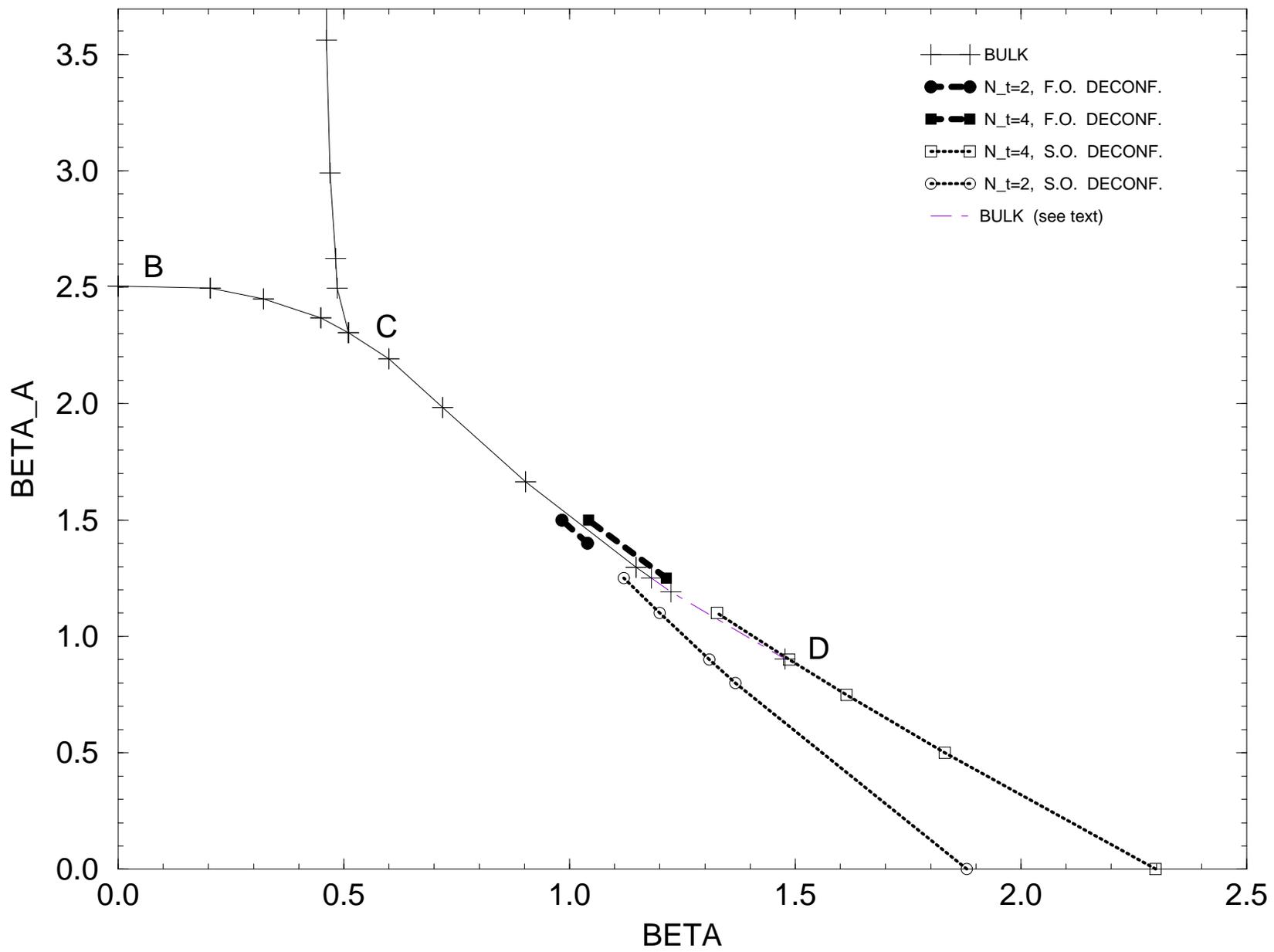
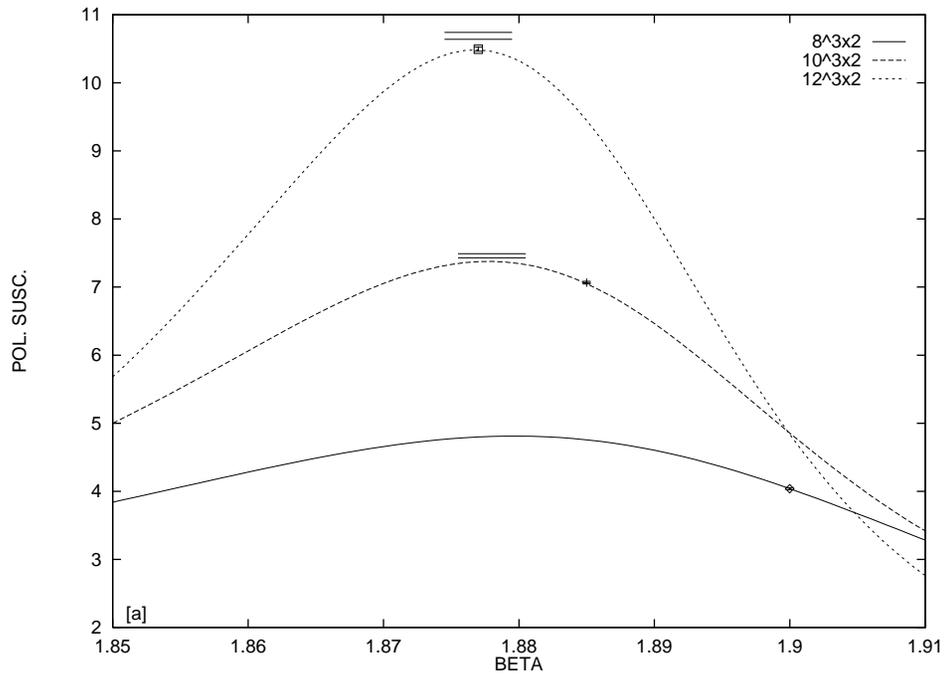
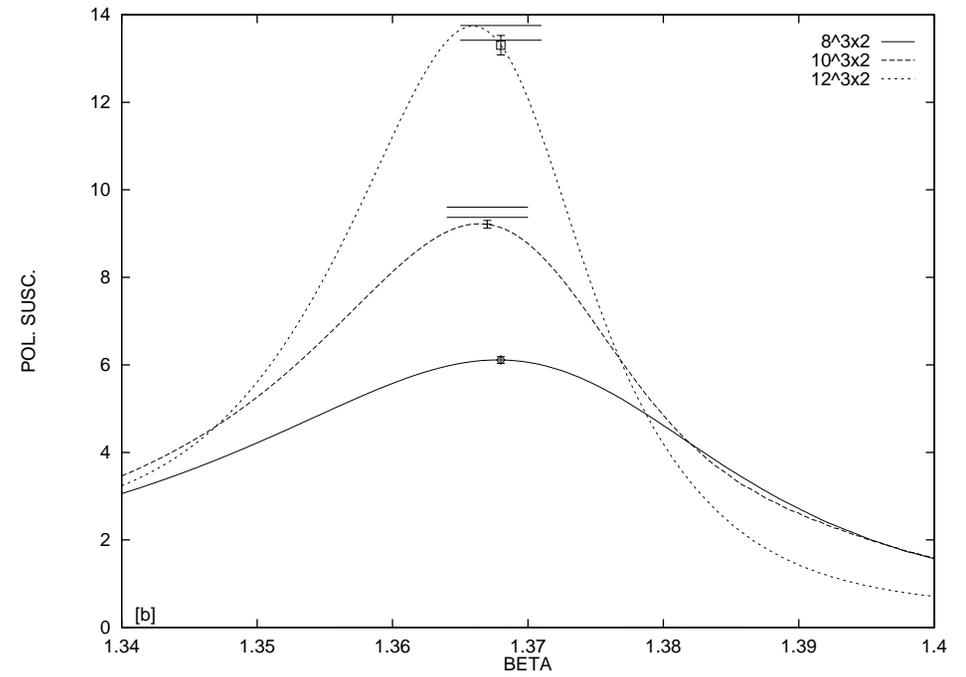
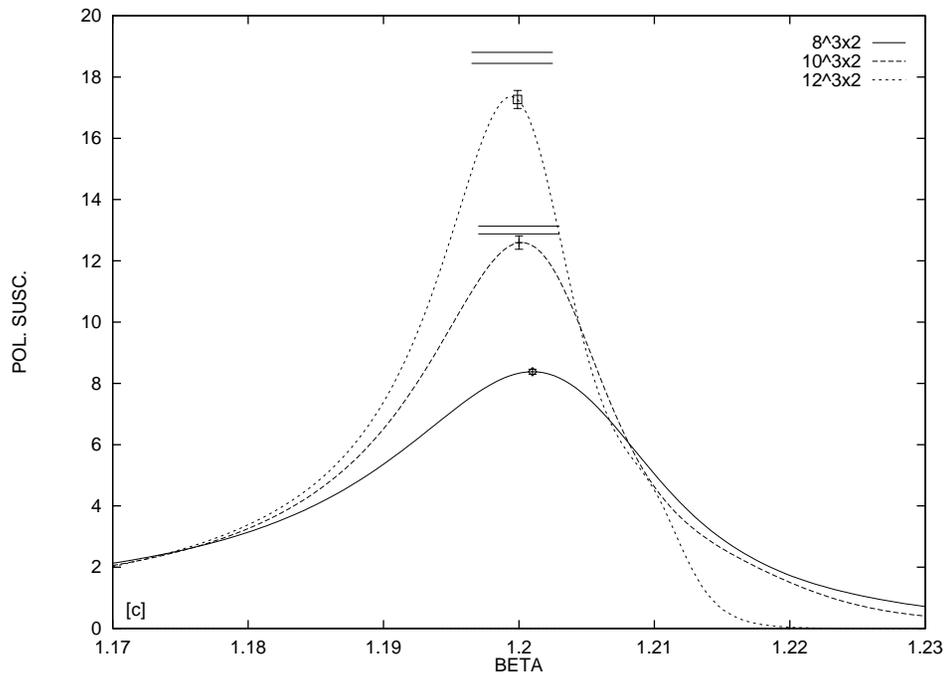
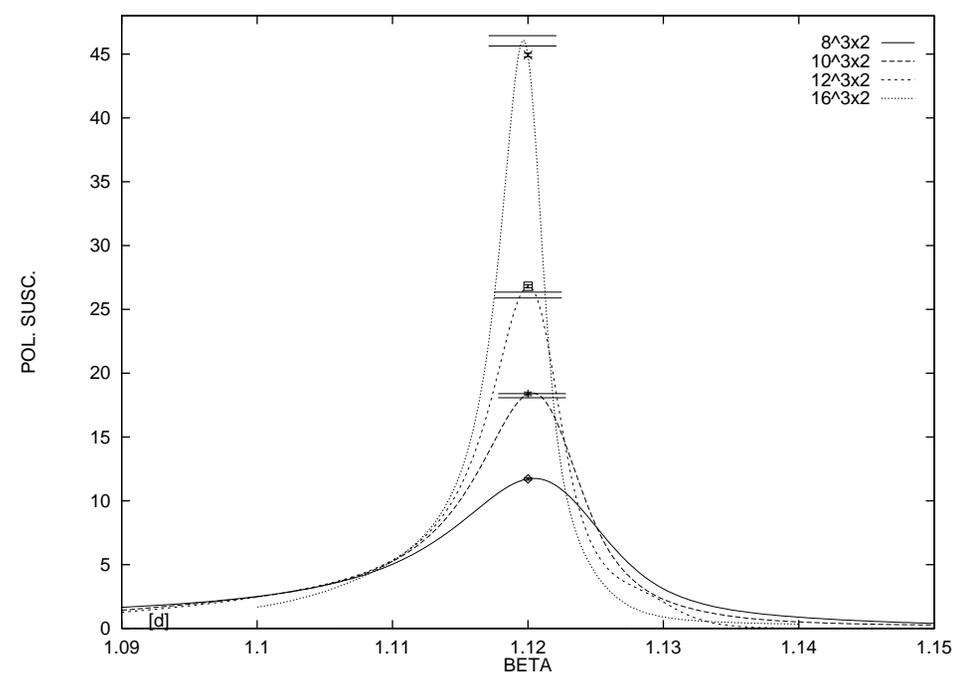
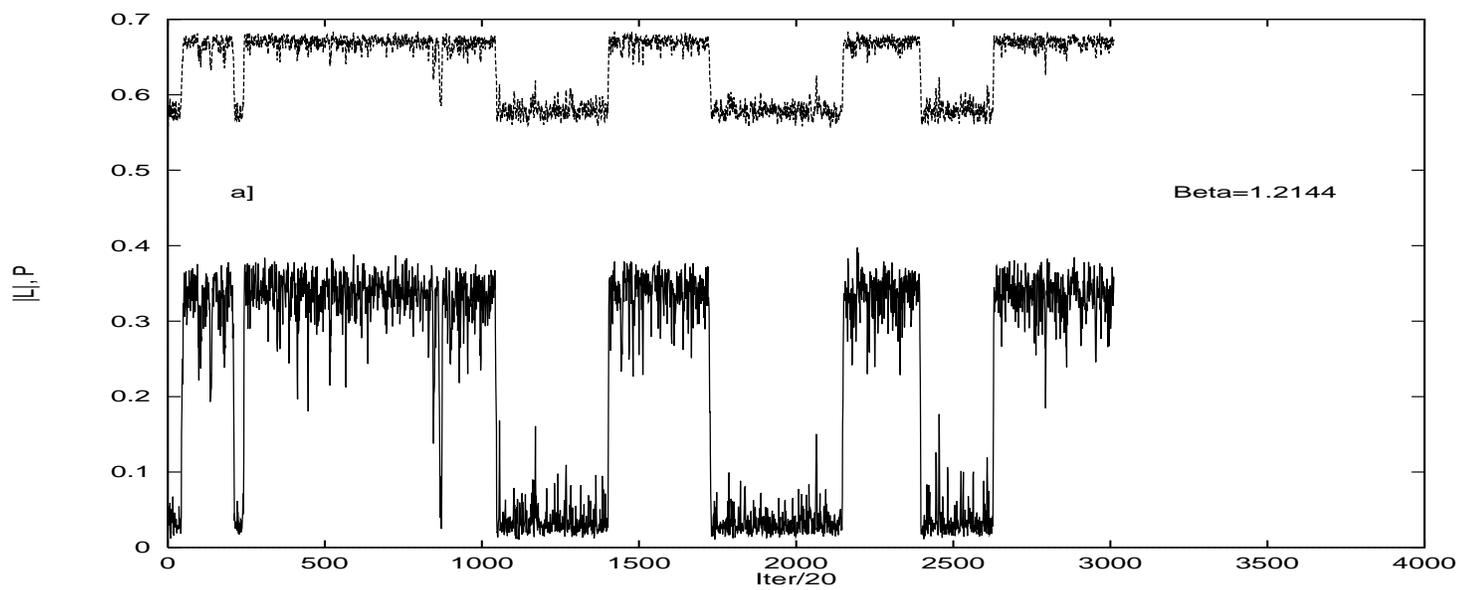
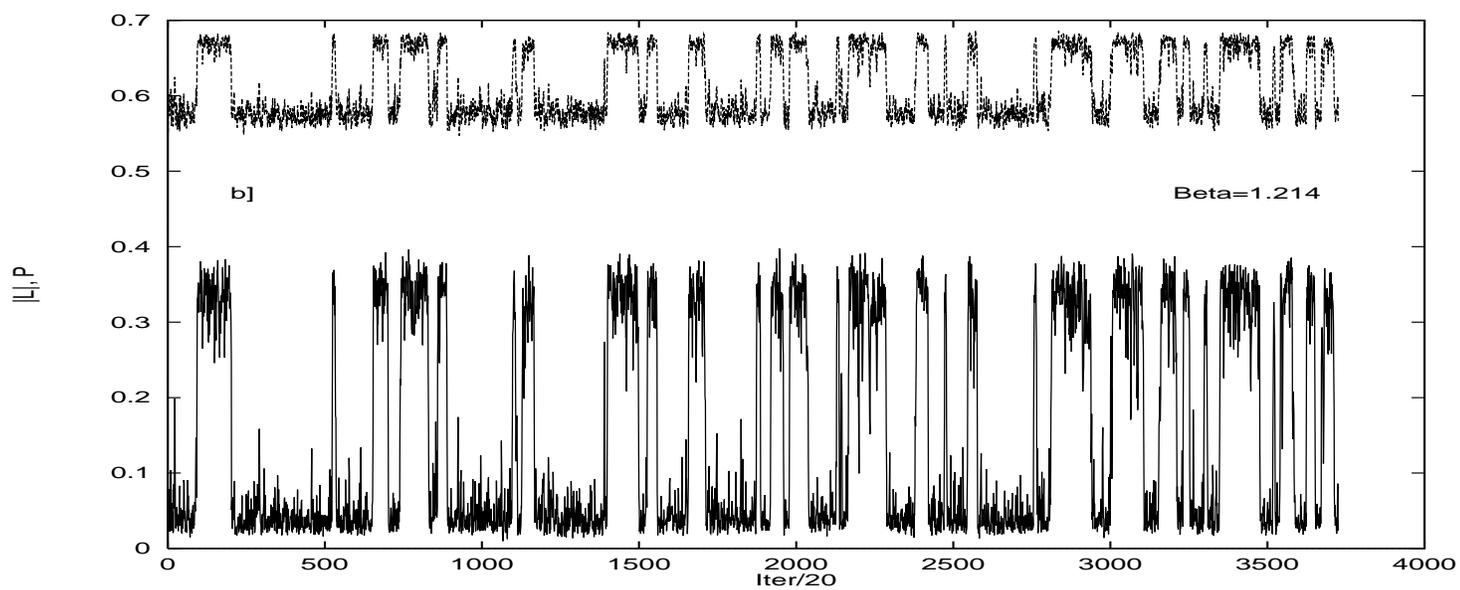
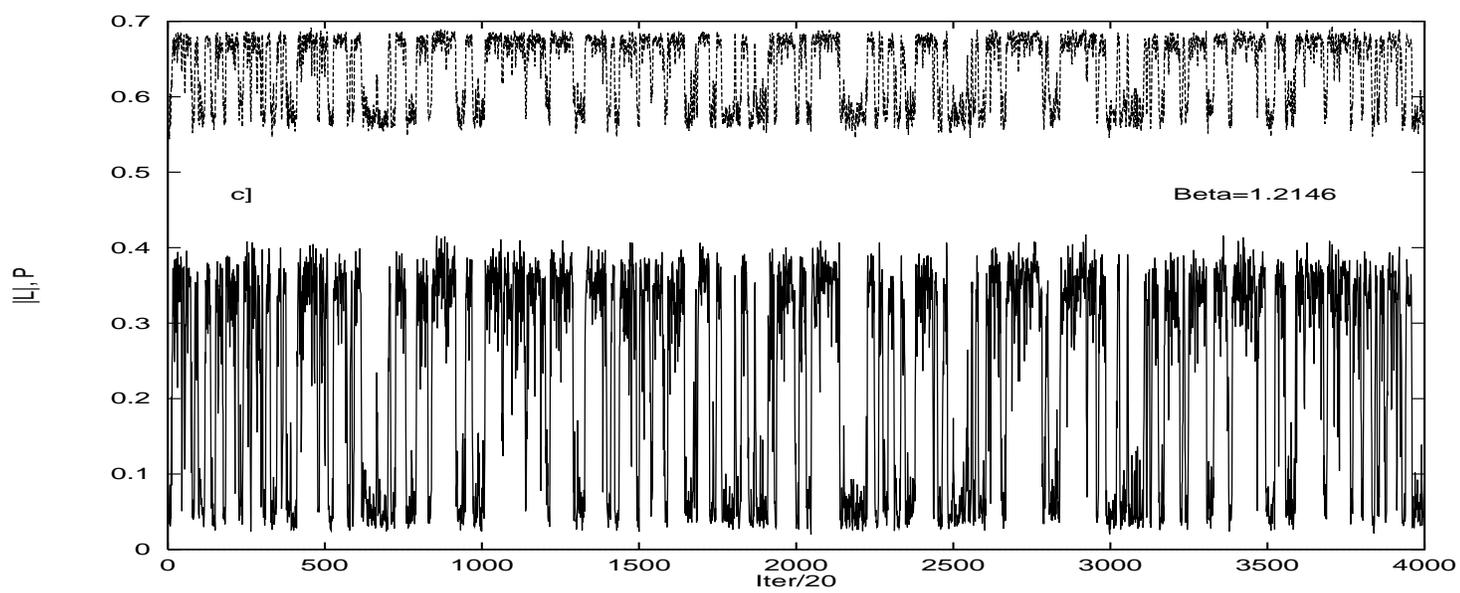
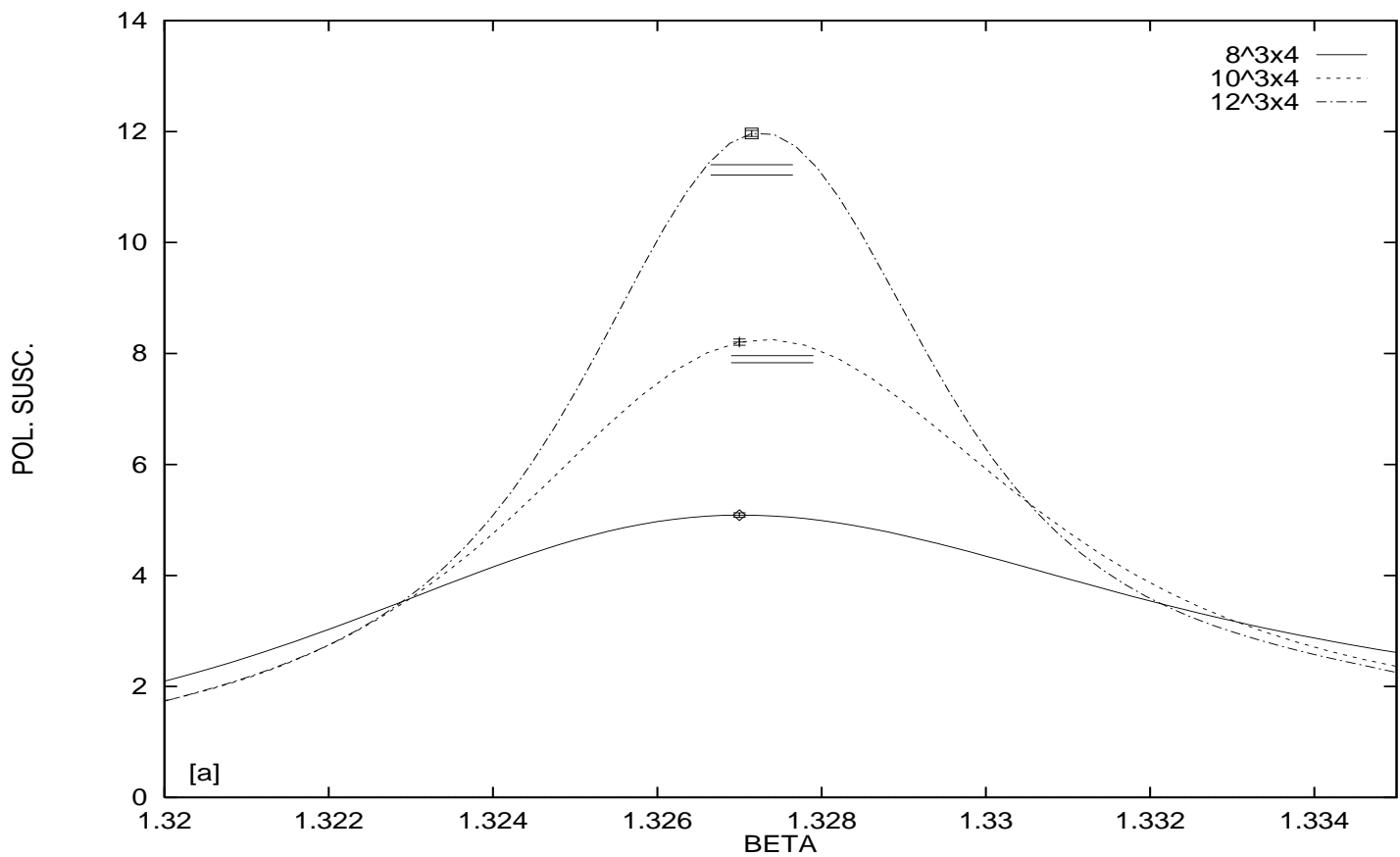
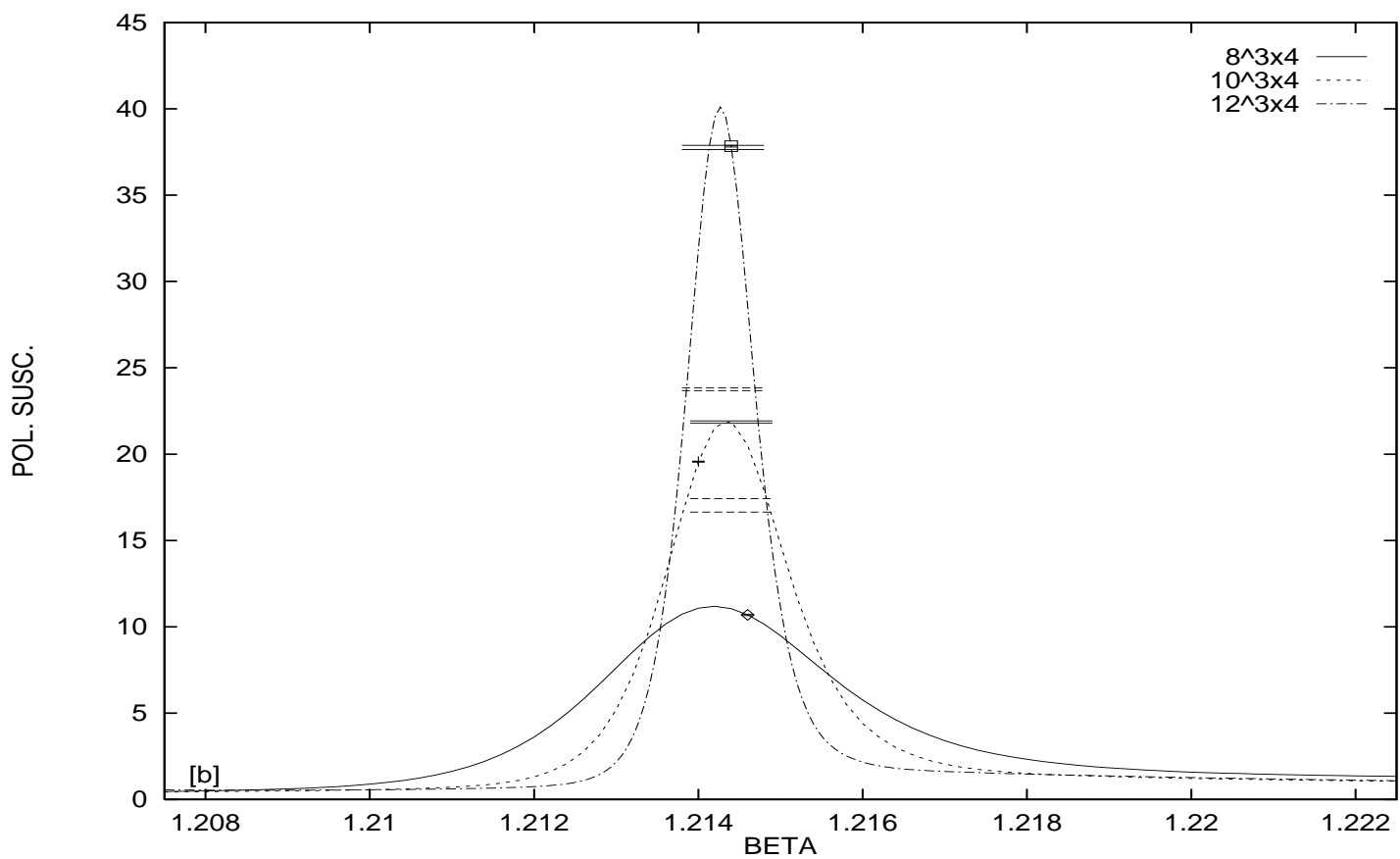
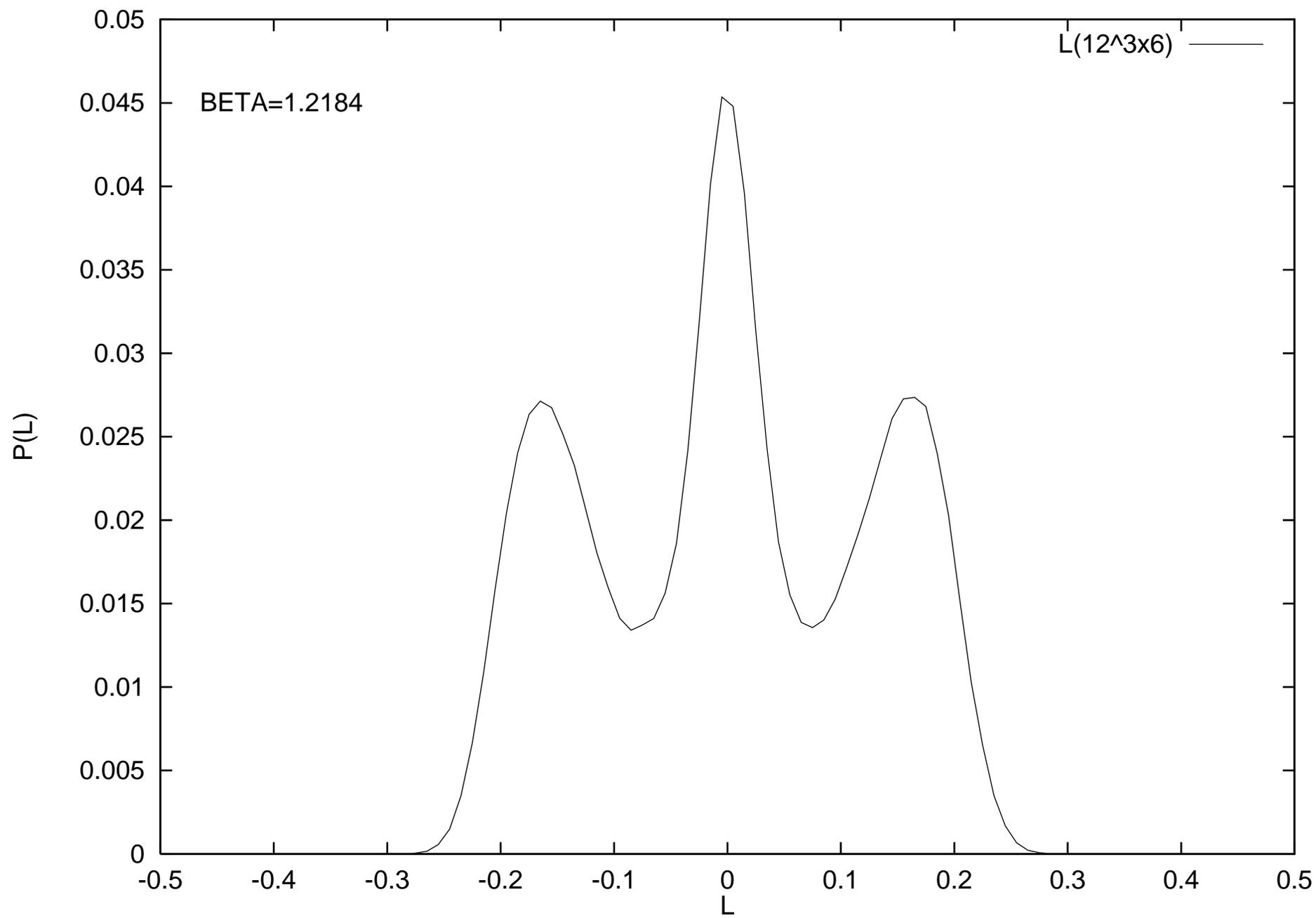
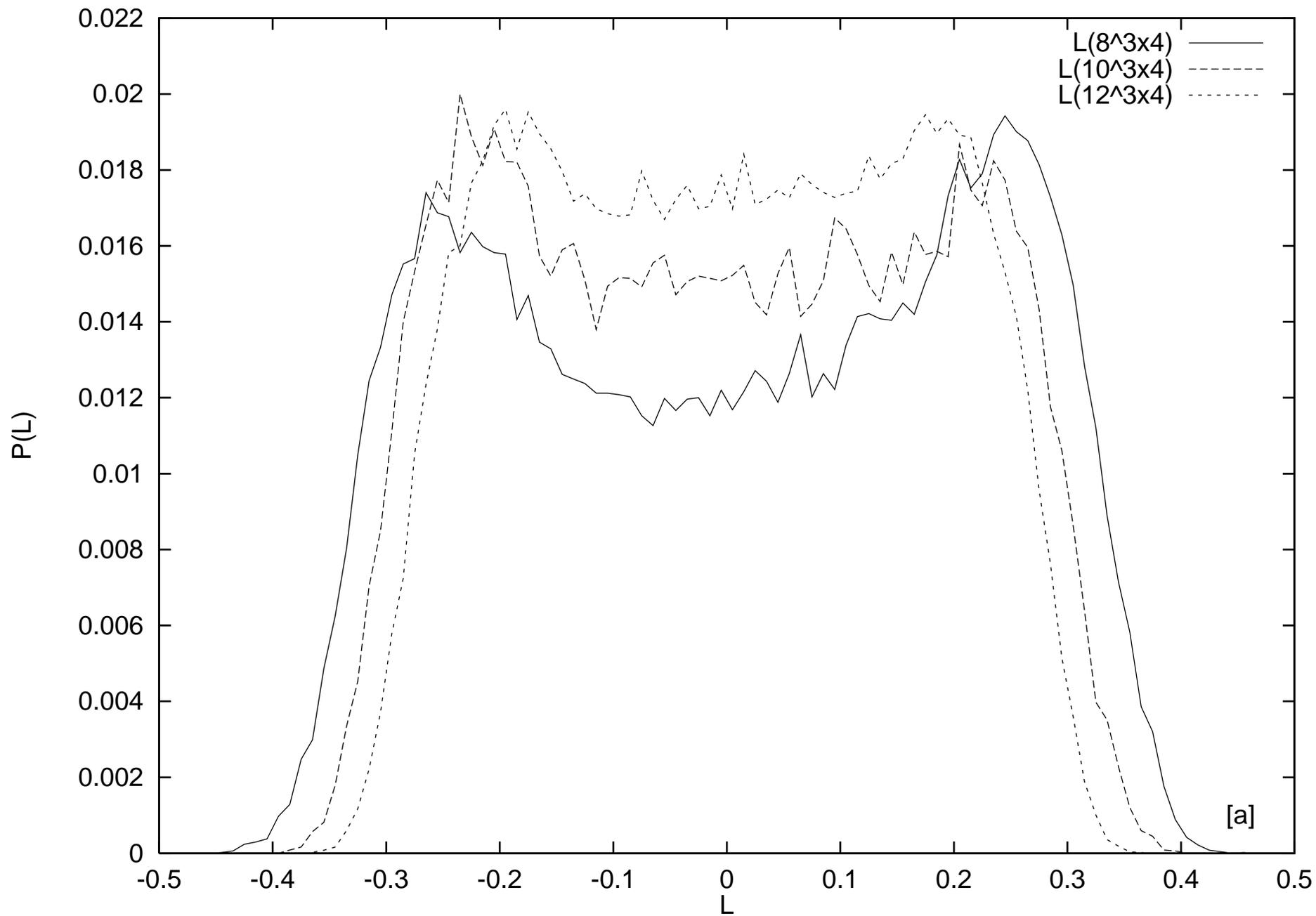
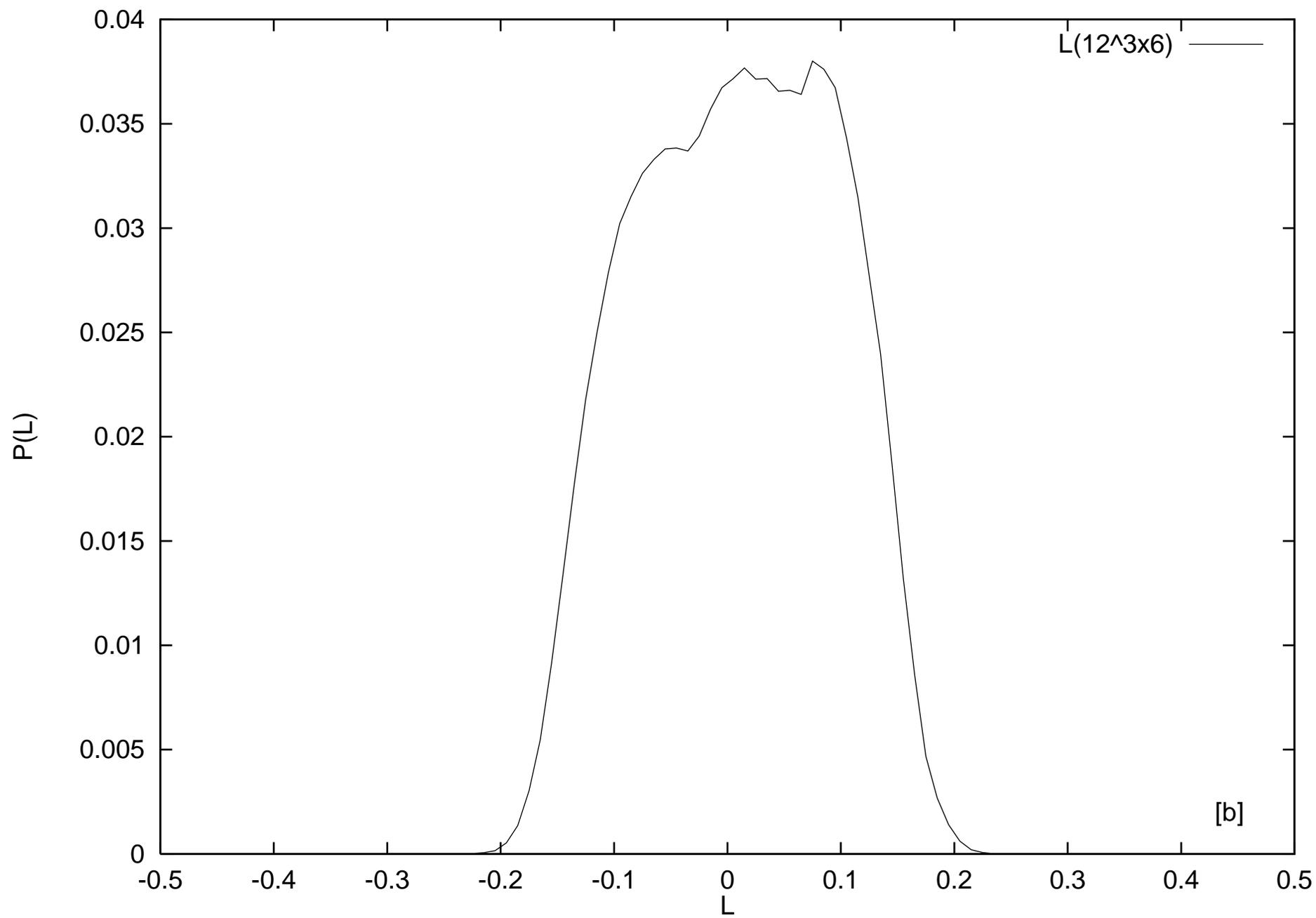
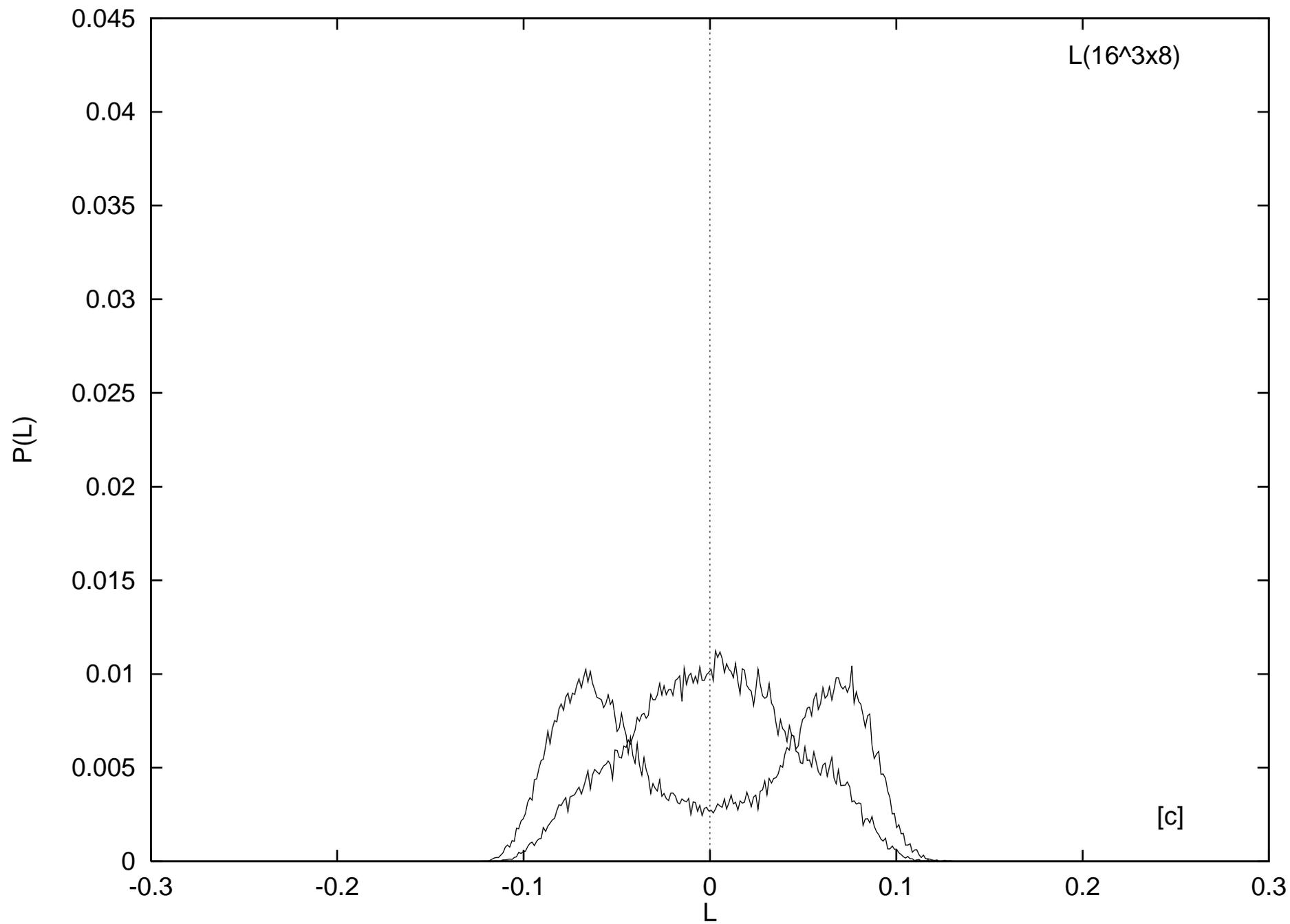
\end{document}